\documentclass[11pt,titlepage,a4paper]{article}
\usepackage{bozhomac}	% loads: full AMSLaTeX and AMSFonts
\usepackage{bozhlogo}	% the logo of BOZHO command \BOZHO
\title{\bfseries		\vspace*{-1.678902345in}
\vspace*{-7ex}
{
\begin{flushright}
	  \textbf{\large LANL xxx E-print archive No. gr-qc/9802057}\\[5ex]
\end{flushright}
}
{\huge On one connection between\\ Lorentzian and Euclidean metrics}
}

\vspace{1.7ex}

\author{
Bozhidar Z. Iliev
\thanks{Department Mathematical Modeling,
Institute for Nuclear Research and \mbox{Nuclear} Energy,
Bulgarian Academy of Sciences,
Boul. Tzarigradsko chauss\'ee~72, 1784 Sofia, Bulgaria}
\thanks{E-mail address: bozho@inrne.bas.bg}
\thanks{URL: http://www.inrne.bas.bg/mathmod/bozhome/}
}

\date{
 \vspace{2.27ex}\ShortTitle{Lorentzian and Euclidean metrics}	\\[0.27ex]
 \vspace{3.27ex}
	\begin{tabular}{r@{$\colon\to~$}l}
 \vspace{0.09ex} Basic ideas	& December 26, 1997	\\[0.09ex]
 \vspace{0.09ex} Began		& December 30, 1997	\\[0.09ex]
 \vspace{0.09ex} Ended		& January 1,  1998 	\\[0.09ex]
 \vspace{0.09ex} Revised	& February, June--July, 1998	\\[0.09ex]
 \vspace{0.09ex} Revised	& January, 1999		\\[0.09ex]
 \vspace{0.09ex} Last update 	& February 16, 1999	\\[0.09ex]
 \vspace{0.27ex} Produced	& \fbox{\today}		\\[0.27ex]
	\end{tabular} \\[1.27ex]
	\begin{tabular}{r@{$\colon~$}l}
 \vspace{0.27ex} LANL xxx archive server E-print No.& gr-qc/9802057\\[0.27ex]
	\end{tabular} \\[-0.27ex]
 \vspace{4.27ex}{\Huge\BOZHO}	\\[4.27ex]
 \vspace{0.27ex}\Subject{Differential geometry, General relativity}\\[2.27ex]
	\begin{tabular}{r@{\hspace{0.512em}}|@{\hspace{0.512em}}l}
 \vspace{0.27ex}	%	\\[0.27ex]
&
 \vspace{0.27ex}	%	\\[0.27ex]
	\end{tabular} \\[1.27ex]
 \vspace{0.27ex}	\\[0.27ex] }

\begin{document}		% BEGINNING OF THE DOCUMENT

\renewcommand{\thefootnote}{\fnsymbol{footnote}} % special footnote symbols
\maketitle				% the title (page) is put here
\renewcommand{\thefootnote}{\arabic{footnote}}   % usual footnote symbols

 \tableofcontents		% the table of contents is put here

%%%%%%%%%%%%%%%%%%%%%%%%%%%%%%%%%%%%%%%%%%%%%%%%%%%%%%%%%%%%%%%%%%%%%%%%%%%%%
%%%%%									%%%%%
%%%%%		actual beginning of the document			%%%%%
%%%%%									%%%%%
%%%%%%%%%%%%%%%%%%%%%%%%%%%%%%%%%%%%%%%%%%%%%%%%%%%%%%%%%%%%%%%%%%%%%%%%%%%%%

\begin{abstract}

We investigate connections between pairs of (pseudo\ndash)Riemannian metrics
whose sum is a (tensor) product of a covector field with itself.
A bijective mapping between the classes of Euclidean and Lorentzian metrics
is constructed as a special result. The existence of such maps on a
differentiable manifold is discussed. Similar relations for metrics
of arbitrary signature  on a manifold are considered. We point the
possibility that any physical theory based on real Lorentzian metric(s) can
be (re)formulated equivalently in terms of real Euclidean metric(s).

\end{abstract}

\section {Introduction}
\label{Sect1}
\setcounter{equation} {0}

	In~\cite{Pestov96} the time is defined as a congruence of lines on a
real differentiable manifold $M$. The vector field $t$ tangent to this
congruence is called \textit{temporal} field. In the work mentioned is
stated that the Maxwell equations on $M$ with an Euclidean metric $e_{ij}$,
$i,j=1,\dots,n:=\dim M$ are derivable from the standard electromagnetic
Lagrangian on $M$ with a (pseudo-)Riemannian metric $g_{ij}=t_it_j-e_{ij}$,
$t_i:=e_{ij}t^j$. In the paper cited $g_{ij}$ is said to be Lorentzian.
Special metrics $g_{ij}$ of this kind, when the norm of $t$ is $2$ (with
respect to both metrics --- see equation~\eref{3.0+2} below), are considered,
\eg in~\cite[sect.~2.6]{Hawking&Ellis}, \cite[p.~219]{Geroch&Horowitz},
and~\cite[p.~148, lemma~36]{O'Neill}.  A slight more general construction of
the kind mentioned can be found in~\cite[pp.~241--242]{Geroch&Horowitz}. For
it, without an investigation, is stated that it is Lorentzian again, which is
not always the case (see Sect.~\ref{Sect3} below).  In the above
constructions $t$ can also be taken to be the gradient vector field of the
global time function~\cite{Hawking-1969,Geroch&Horowitz}.

	The purpose of the present work is to be investigated pairs of
(pseu\-do-)Riemannian metrics  $(g_{ij},h_{ij})$ whose sum is a product
of the covariant components of some vector field $t$, \ie
$g_{ij}+h_{ij}=t_i t_j$ with, e.g., $t_i:=g_{ij}t^j$.%
\footnote{%
Bundle decompositions and correspondences between various types of metric
tensors are consequences of the Witt (decomposition)
theorem~\cite[chapter~XIV, \S~5]{Lang-algebra}. The present paper deals with
one specific such correspondence based on the use of a vector field $t$ with
appropriate properties.%
}
In particular, we prove
the important for the physics result that for any real Euclidean (resp.
Lorentzian) metric there exists real Lorentzian (resp. Euclidean) metric
forming with it such a pair.

	In Sect.~\ref{Sect2} we prove that if  $g_{ij}$ is an Euclidean
metric, then (for $g_{ij}t^it^j\not=1$) the metric $h_{ij}=t_it_j-g_{ij}$
can be only Lorentzian or negatively definite. As a corollary, we construct a
map from the set of Euclidean metrics into the set of Lorentzian ones.
To the applicability of the results of Sect.~\ref{Sect2} is devoted
Sect.~\ref{applicability}. Here we point to some topological obstacles that
may arise in this direction and formulate these results in a form of a
proposition.
The general case, for arbitrary (pseudo-)Riemannian metric $g_{ij}$, is
investigated in Sect.~\ref{Sect3}. If $g_{ij}$ has a signature $(p,q)$, \ie
if the matrix $[g_{ij}]$ has $p$ positive and $q$ negative eigenvalues,%
\footnote{%
Some times the pair $(p,q)$ is called type of $g$ and the signature is
defined as the number $s=p-q$. In this paper we suppose the numbers $p$ and
$q$  to be independent of the point at which they are calculated, \ie here we
consider metrics whose signature is point-independent and so constant over
the corresponding sets. The numbers $p$ and $q$ are also known as positive
index and (negative) index of the metric. Often, especially in the physical
literature, the signature is defined as an order $n$\ndash tuple
$(\varepsilon_1,\ldots,\varepsilon_n)$
where $p$ (resp.\ $q=n-p$) of $\varepsilon_1,\ldots,\varepsilon_n$ are equal
to $+1$ (resp.\ $-1$) or simply to the plus (resp.\ minus) sign and the order
of $\varepsilon_1,\ldots,\varepsilon_n$ corresponds to the one of the signs
of the diagonal elements of the metric in some (pseudo\ndash)orthogonal basis%
}
then the signature of $h_{ij}$, if
it is non-degenerate, which is the conventional case, can be
$(q,p)$ or $(q+1,p-1)$. As a side-result, we obtain a map from the set of all
Lorentzian metrics into the set of Euclidean ones. Some inferences of the
results obtained are presented in Sect.~\ref{Sect4}. We construct
bijective maps from the set of metrics with signature $(p,q)$ on that with
signature $(q+1,p-1)$, which, in particular, is valid for the classes of
Euclidean and Lorentzian metrics.%
\footnote{%
A (partial) correspondence between Euclidean and Lorentzian metrics is
established in~\cite{Barbero96} via the Einstein equations.
}
We also correct some wrong statements
of~\cite{Pestov96}. Some concluding remarks are presented in
Sect.~\ref{conclusion}. In particular, we construct, possibly under some
conditions, bijective real maps between (pseudo-)Riemannian metrics of
arbitrary signature.
% In the Appendix are separated some technical calculations.

	Now, to fix the terminology, which significantly differs in
different works, we present some definitions.

	Following~\cite[p.~273]{Bruhat}, we call Riemannian metric on a real
manifold $M$ a non-degenerate, symmetric and 2-covariant tensor field $g$ on
it. If for any non-zero vector $v$ at $x\in M$ is fulfilled $g_x(v,v)>0$, the
metric is called proper Riemannian, positive definite, or Euclidean;
otherwise it is called indefinite or pseudo-Riemannian~\cite{Bruhat,K&N-1}.
It is known that every finite-dimensional paracompact differentiable manifold
admits positively definite (Euclidean)
metrics~\cite[chapter~IV, \S~1; chapter~I, example~5.7]{K&N-1},
	\cite[chapter~1, excercise~2.3]{Warner}, \cite[p.~280]{Bruhat}.
A pseudo-Riemannian metric with exactly one positive eigenvalue is called
Lorentzian~\cite{Hawking&Ellis} (or, some times, Minkowskian).%
\footnote{%
One can also find the definition of a Lorentzian metric as a metric with only
one negative eigenvalue~\cite[p.~55]{O'Neill}. This definition is isomorphic
to the one used in the present paper (see,
e.g.,~\cite[pp.92\Ndash93]{O'Neill}.%
}
If in the above definitions the non-degeneracy condition is dropped, the
prefix `semi-' is added to the names of the corresponding
metrics~\cite{Rosenfel'd}; e.g.\  a semi-Riemannian metric on $M$ is a
symmetric two times covariant tensor field on it~\cite{Rosenfel'd}.

\section {Euclidean case}
\label{Sect2}
\setcounter{equation} {0}

	Let $e$ be an Euclidean metric on a finite-dimensional,
paracompact, differentiable, and real manifold $M$ and $t$ a vector field on
$M$. Consider the 2-covariant symmetric tensor field
	\begin{equation}	\label{2.1}
g=e(\cdot,t)\otimes e(\cdot,t) - e ,
	\end{equation}
where $\otimes$ is the tensor product sign. In a local chart its
local components are
	\begin{equation}	\label{2.2}
g_{ij}=t_i t_j - e_{ij},\qquad t_i=e_{ij}t^j.
	\end{equation}
Here and below the Latin indices run form 1 to $n:=\dim M<\infty$ and a
summation from 1 to $n$ over indices repeated on different levels is assumed.

% New proof: Jan 4--11, 1999 ++++++++++++++++++++++++++++++++++++++++++++++++

	Let $x$ be arbitrary point in $M$.

	If $t|_x=0$, then $g_{ij}=-e_{ij}$ and, consequently,  $g$ is at $x$
a Riemannian metric with signature $(0,n)$ as that of $e$ is $(n,0)$.

	If $t|_x\not=0$, then we can construct by means of a
Gramm\ndash{Schmidt}
procedure~\cite[chapter~4, \S~3]{Bellman}, \cite[pp.~206--208]{Kurosh}
an orthonormal (with respect to $e$) basis $\{E_i\}$ in the tangent to $M$
space at $x$ with
$E_1=\frac{1}{\alpha}t$ where $\alpha=+\sqrt{e(t,t)}>0$.
With respect to such a basis we have
 $e_{ij}=\delta_{ij}$,
 $t^i=\alpha\delta_{1}^{i}$, and
 $t_i=e_{ij}t^j=\alpha\delta_{i1}$
where $\delta_{ij}:=\delta_{i}^{j}:=\delta^{ij}$ are the Kroneker symbols,
\ie $\delta_{ij}=1$ for $i=j$ and $\delta_{ij}=0$ for $i\not=j$.
In the basis mentioned the matrix of $g$ is diagonal and
	\begin{equation}	\label{2.3}
[g_{ij}] = \diag \bigl(e(t,t)-1,-1,\dots,-1\bigr)
	\end{equation}
due to~\eref{2.2}. Therefore $g$ is a Riemannian metric on $M$ if and only
if  the Euclidean norm of $t$ is not equal to one at every point, \ie
	\begin{equation}	\label{2.4}
e(t,t)\not=1 .
	\end{equation}
From~\eref{2.3} we see that the first eigenvalue of $g$ is
 $\lambda_1\gtreqless0$
iff
 $e(t,t)|_x\gtreqless1$,
the remaining $n-1$ eigenvalues being equal to $-1$.
Since the metric's signature
is independent of the local coordinates by means of which it is computed,%
\footnote{%
This is a trivial corollary of the results
of~\cite[chapter~XIV, \S7]{Lang-algebra}.%
}
from here an important result follows:
if $e(t,t)>1$ at every point, the metric $g$ is Lorentzian,
if $e(t,t)<1$ at every point, it is negatively definite and, hence,
					isomorphic to an Euclidean one,
and for $e(t,t)=1$ at every point it is semi\ndash{Riemannian} with signature
$(0,n-1,1)$.

	Summing up, if there exists a vector field $t$ satisfying~\eref{2.4}
or $e(t,t)=1$ at every point, then~\eref{2.1} defines a metric on $M$ for
which there are three possibilities: First, if $e(t,t)>1$, it is Lorentzian.
Second, if $e(t,t)=1$, it is semi-Riemannian, viz.\ a 1-degenerate metric,
and, consequently, non-Riemannian one~\cite{Rosenfel'd}.  And third, if
$e(t,t)<1$, it is negatively definite, and so isomorphic to an Euclidean
metric. From physical view-point, the most essential result is that
if for every $e$ we choose some vector field $t_e$ with $e(t_e,t_e)>1$, then
the mapping $e\mapsto g$, given by~\eref{2.1} for $t=t_e$,
maps the class of Euclidean metrics on $M$ into the class of Lorentzian ones.
It is clear, this map essentially depends on the choice of the vector
fields $t_e$ used in its construction.

\section {Applicability of the results}
\label{applicability}

	Up to this point we have supposed two major things: the existence of
an Euclidean metric $e$ and of a vector field $t$ with the corresponding
properties on the \emph{whole} manifold $M$. In this sense the above
considerations are global. Of course, we can localize them by restricting
ourselves on some subset $U$ of $M$. Different conditions for global or
local existence of (Euclidean) metrics are well-known and discuss at length
in the corresponding literature (see, e.g.,~\cite[chapter~IV]{Lang-manifolds}
or~\cite{K&N-1,Greub&et.al.-1}). In our case, the existence of Euclidean
metric on $M$ is a consequence of the paracompactness and finite\ndash
dimensionality of the manifold $M$~\cite{K&N-1}. These assumptions are enough
for the most physical applications and we assume they are valid in this
work.%
\footnote{%
See the partial discussion of this problem
in~\cite[sect.~5.2]{Geroch&Horowitz}.%
}

	What concerns the existence of a vector field $t$ with properties
required ($e(t,t)$ to be greater than, or equal to, or less than one), some
problems may arise. If on $t$ we do not impose additional restrictions, it
always can be constructed as follows: Take a non\ndash vanishing on $M$ vector
field $t_0$,%
\footnote{%
Generally $t_0$ is discontinuous (see below).%
}
so $e(t_0,t_0)\not=0$ (everywhere on $M$). Defining
$t:=\sqrt{a}t_0/\sqrt{e(t_0,t_0)}$ for $a\in\mathbb{R}$, $a\ge0$, we get
$e(t,t)=a$.  Hence, choosing $a\gtreqqless 1$, we obtain $e(t,t)\gtreqqless
1$. Obviously, the existence of $t$ in the first two cases, $e(t,t)\geq1$, is
equivalent to the existence of a non\ndash vanishing vector field on $M$,
while in the last one, $e(t,t)<1$, this is not necessary, viz.\ in it $t$ may
vanish on some subsets on $M$ or even to be the null vector field on $M$.

	The general conclusion is: a vector field the with $e(t,t)\geq1$
(over $M$) exits iff $M$ admits nowhere vanishing (on $M$) vector field. Thus,
our results concerning the case $e(t,t)\geq1$ are applicable iff such a field
exists. As we said above, this is just the situation if we do not impose
additional conditions on $t$. But this is not satisfactory from the
view-point of concrete applications. For instance, in the most
mathematical investigations the (Euclidean or (semi-)Riemannian) metrics are
required to be differentiable of class
$C^1$~\cite{K&N-1,Greub&et.al.-1,Greub&et.al.-2,Bruhat}; \eg in the Riemannian
geometry one normally uses $C^2$ metrics. Such an assumption implies $t$ to
be of class of smoothness at least $C^1$. Analogous is the situation in
physics, for example, the treatment of $t$ as a temporal field requires $t$
to be at least continuous~\cite{Pestov96} and the considerations on the
background of general relativity force us to assume $t$ to be of class
$C^2$~\cite{Hawking&Ellis}.

	Therefore of great importance is the case when the vector field $t$
satisfies certain smoothness conditions, viz.\ when it is of class $C^m$ for
some $m\geq0$. At this point some topological obstacles may arise for the
global existence of $t$ with $e(t,t)\geq1$. In fact, the above-said implies
that a vector field $t$ of class $C^m$ with $e(t,t)\geq1$ exists on
$U\subseteq M$ iff on $U$ there exists a $C^m$ non-vanishing vector field.
But it is well-known that not every manifold admits such a tangent vector
field~\cite{Spivak-1}.
A classical example of this kind are the even\ndash dimensional spheres
$\mathbb{S}^{2k}$, $k\in\mathbb{N}$:
on $\mathbb{S}^{2k}$ does not exist non-vanishing
(on the whole $\mathbb{S}^{2k}$) continuous vector
field~\cite{Spivak-1},~\cite[sect.~4.24]{Schutz}.
Examples of the opposite kind are the odd\ndash dimensional spheres
$\mathbb{S}^{2k-1}$~\cite{Spivak-1},~\cite[excerise.~4.26]{Schutz}
and the path\ndash connected manifolds with flat $C^1$ linear connection:
they always admits global $C^1$ non-vanishing vector fields.%
\footnote{%
In the last case such a vector field can be constructed as follows. Fix a
non-zero vector $v_0$ at an arbitrary point $x_0\in M$.  Define the vector
field $v$ at any $x\in M$ as the result of the parallel transport, assigned
to the given flat connection, of $v_0$ from $x_0$ to $x$ along some path
connecting $x_0$ and $x$.  Then $v$ is a tangent vector field on $M$ which is
non-vanishing and of class $C^1$.%
}
Also every non\ndash compact manifold admits $C^0$ non\ndash zero vector
field~\cite{Markus}. An analysis of the question of existence of vector
fields (and Lorentz metrics) can be found in~\cite{O'Neill} where also other
examples are presented. Consequently, the global existence of $C^m$, $m\geq0$
field $t$ with $e(t,t)\ge1$ depends on the concrete manifold $M$ and has to
be investigated separately for any particular case.

	In conclusion, the results of the preceding section are valid
locally and for their global, \ie on the whole manifold $M$, validity may
arise obstacles of pure topological nature. Since on $M$,  due to the
paracompactness and finite dimensionality, an Euclidean metric always
exists, this is connected with the existence of a vector field $t$ with
properties required. So, the precise formulation of the results obtained is
the following.

	\begin{Prop}	\label{Euclidean}
	Let $t$ be a vector field over $U\subseteq M$, $e$ be an Euclidean
metric on $U$, and
$U_{\gtreqqless}:=\{x\; |\quad x\in U,\ e(t,t)|_x\gtreqqless 1\}$.
Then the tensor field~\eref{2.1} is:

\textup{\textbf{\hphantom{ii}(i)}}
a negatively definite Riemannian metric on $U_<$;

\textup{\textbf{\hphantom{i}(ii)}}
a Lorentzian metric on $U_>$;

\textup{\textbf{(iii)}}
a 1-degenerate negatively definite semi-Riemannian metric on $U_=$.
	\end{Prop}

	The most interesting is the `smooth' global case when $U=M$, two of
the sets $U_<$, $U_>$, and $U_=$ are empty, and $t$ is of class $C^m$,
$m\ge0$. As we notice above, one can always choose $t$ such that
$U_>=U_==\varnothing$ and $U_<=M$ but the two other cases, $U_>=M$
and~$U_==M$,  can not be realized for arbitrary manifold~$M$ if~$t$ is of
class $C^m$ with $m\ge0$. If we drop the smoothness requirement, then $t$ can
always be chosen such that one of the sets $U_<$, $U_>$, and $U_=$ to be
equal to $M$, the other two being the empty set.

\section {General case}
\label{Sect3}
\setcounter{equation}{0}

	It is said that a Riemannian metric $g$ on  $U\subseteq M$ is of
signature $(p,q)$, $p+q=n:=\dim M$, if it has $p$ positive and $q$ negative
eigenvalues. A semi-Riemannian metric on $U$ is of signature $(p,q)$ and
defect $r$ (or of signature $(p,q,r)$, or  $r$\ndash{degenerate} with
signature $(p,q)$), $p+q+r=n$, if it has $p$ positive, $q$ negative, and
$r$ vanishing eigenvalues.

	\begin{Prop}	\label{Prop3.1}
Let $g$ be a Riemannian metric of signature $(p,q)$ on $U\subseteq M$, $t$ be
a vector field on $U$,
$\widetilde{U}_{\gtreqqless}^{+}:=\{x|\ x\in U,\ g(t,t)|_x\gtreqqless 1\}$,
and
	\begin{equation}	\label{3.1}
g\mapsto\Tilde{g}^+ := h := g(\cdot,t)\otimes g(\cdot,t) - g.
	\end{equation}

Then the tensor field $h$ is:

\textup{\textbf{\hphantom{ii}(i)}}
a Riemannian metric with signature $(q,p)$ on $\widetilde{U}_{<}^{+}$.

\textup{\textbf{\hphantom{i}(ii)}}
a Riemannian metric with  signature $(q+1,p-1)$ on $\widetilde{U}_{>}^{+}$.

\textup{\textbf{(iii)}}
a (parabolic) semi-Riemannian metric with signature $(q,p-1)$
and defect $1$ on $\widetilde{U}_{=}^{+}$, \ie on $\widetilde{U}_{=}^{+}$
the bilinear map $h$ has $q$ positive, $(p-1)$ negative, and 1 vanishing
eigenvalue.
	\end{Prop}

	\begin{Proof}
Since $g$ is by definition 2-covariant symmetric tensor field, such is $h$
too. So, it remains to be studied the eigenvalues of $h$.

	Let $x\in U$ be arbitrarily fixed point. We shall prove the
proposition at $x$, \ie for $U=\{x\}\subset M$. Then the general result will
be evident as $U=\bigcup_{x\in U}\{x\}$. All of the bellow written quantities
in this proof will be taken at $x$; so their restriction at $x$ will not be
written explicitly. We shall distinguish three cases.

	`Null' case, $t=0$. The statement is evident as now $g(t,t)=0<1$ and
$h=-g$; so the signature of $h$ is $(q,p)$.

	`Non-isotropic' case, $t$ is non-isotropic, \ie  $g(t,t)\not=0$ and
hence $t\not=0$. Let $\{E_{i}^{\prime}\}$ be a basis in $T_x(M)$, the space
tangent to $M$ at $x$, and consisting of non-isotropic vectors with $E_1=t$.
Applying to this basis the standard Gramm\ndash{Schmidt} orthogonalization
procedure~\cite[chapter~4, \S~3]{Bellman}, \cite[pp.~206--208]{Kurosh},
with respect to the scalar product $(\cdot,\cdot)=g(\cdot,\cdot)$, we can
construct (after normalization) a (pseudo\ndash)Riemannian basis $\{E_i\}$ (at
$x$) such that $E_1=t/\alpha$, $\alpha:=+\sqrt{|g(t,t)|}$ and
$g_{ij}:=g(E_i,E_j)=\varepsilon_i\delta_{ij}$ ($i$ is not a summation index
here!) where $p\in\mathbb{N}\bigcup\{0\}$ of the numbers
$\varepsilon_1,\dots,\varepsilon_n$ are equal to $+1$ while the others
$q=n-p$ of them are equal to $-1$ and $\delta_{ij}$ are the Kroneker deltas.
With respect to $\{E_i\}$, we easily obtain
	\begin{equation}	\label{3.h}
[h_{ij}] = \diag
	\big(\varepsilon_1(g(t,t)-1),-\varepsilon_2,\dots,-\varepsilon_n\big),
\qquad
\varepsilon_1=\sign(g(t,t)) .
	\end{equation}
From here the formulated results follow immediately.

	`General isotropic' case, $t$ is non-zero and isotropic, \ie
$t\not=0$ and $g(t,t)=0$. As it is easily seen, this is possible only for
$n:=\dim M\geq2$ and $q,p\geq1$. Now the above method can not be applied
directly because the Gramm\ndash{Schmidt} procedure fails if some of the
initial vectors are/is isotropic (cf., e.g., its construction
in~\cite[pp.~206--208]{Kurosh}). Let $\{E_{i}^{\prime\prime}\}$ be some fixed
basis in which the components of g are
$g_{ij}^{\prime\prime}=\varepsilon_{i}^{\prime\prime}\delta_{ij}$,
$\varepsilon_{i}^{\prime\prime}=\pm1$.
Since $t\not=0$ and
$0=g(t,t)=\sum_i\varepsilon_{i}^{\prime\prime}(t^{\prime\prime\,i})^2$,
at least two of the components  $t^{\prime\prime\,i}$ of $t$ in
$\{E_{i}^{\prime\prime}\}$ are non\ndash{null}. Let these two non\ndash{zero}
components correspond to $i=1,2$, which can be achieved by an appropriate
renumbering of the initial basis.
Let $\{E_{i}^{\prime}\}$ be a basis of non\ndash{isotropic} vectors such
that
$E_{1}^{\prime}$ and $E_{2}^{\prime}$ are fixed and in
$\{E_{i}^{\prime\prime}\}$ their components are
\[
(E_{k}^{\prime})^{\prime\prime\,1} =
	\frac{1}{a}\, t^{\prime\prime\,1}\, \delta_{k1}, \qquad
(E_{k}^{\prime})^{\prime\prime\,i} =
	\frac{1}{a}\, t^{\prime\prime\,i}\, \delta_{k2}, \qquad
\text{for $k=1,2$ and $i\geq2$},
\]
where
\(
a := +\sqrt{| \varepsilon_{1}^{\prime\prime}(t^{\prime\prime\, 1})^2|}
   = |t^{\prime\prime\, 1}|
   = +
	\left|
	\sum_{i\geq2} \varepsilon_{i}^{\prime\prime} (t^{\prime\prime\, i})^2
	\right|^{{1}/{2}}
   > 0
\)
due to
\(
0=g(t,t)=
	\sum_{i} \varepsilon_{i}^{\prime\prime} (t^{\prime\prime\, i})^2
\)
and
 $t^{\prime\prime\, 1}, t^{\prime\prime\, 2}\not=0$.
A simple computation in $\{E_{i}^{\prime\prime}\}$ shows that
\(
t^{\prime\prime\,i} =
  a\bigl(
	(E_{1}^{\prime})^{\prime\prime\,i} +
				(E_{2}^{\prime})^{\prime\prime\,i}
\bigr)
\)
and
 $g(E_{i}^{\prime},E_{j}^{\prime})=\varepsilon_{i}^{\prime\prime}\delta_{ij}$
for $i,j=1,2$. So $E_{1}^{\prime}$ and $E_{2}^{\prime}$ are
non\ndash{isotropic} and
 $t=a(E_{1}^{\prime}+E_{2}^{\prime})$.
Now applying to $\{E_{i}^{\prime}\}$ the Gramm\ndash{Schmidt} procedure (for
the scalar product $(\cdot,\cdot)=g(\cdot,\cdot)$) and normalizing the
vectors, we get a (pseudo\ndash)orthogonal basis $\{E_i\}$ such that
\[
g_{ij}:=g(E_i,E_j)=\varepsilon_i\delta_{ij},
\qquad
E_1=E_{1}^{\prime},
\quad
E_2=E_{2}^{\prime},
\quad\text{and}\quad
t=a(E_1+E_2)
\]
where $p$ of $\varepsilon_1,\dots,\varepsilon_n$ are equal to $+1$ and
$q=n-p$ of them are equal to $-1$
(and also $\varepsilon_1=\varepsilon_{1}^{\prime\prime}$,
 $\varepsilon_2=\varepsilon_{2}^{\prime\prime}$).
Since in $\{E_i\}$ the components of $t$ are
 $t^i=a(\delta^{i1}+\delta^{i2})$, we have
 $t_i=g_{ij}t^j=a\varepsilon_i(\delta_{i1}+\delta_{i2})$.
The equality  $\varepsilon_1+\varepsilon_2=0$ is also valid by virtue of
 $g(t,t)=0$ and $a\not=0$.
Therefore in $\{E_i\}$ the matrix of $h$ is
	\begin{equation}	\label{3.h-isotropic}
[h_{ij}] =
\diag\left(
	\begin{pmatrix}
a^2-\varepsilon_1 		& a^2\varepsilon_1\varepsilon_2	\\
a^2\varepsilon_1\varepsilon_2	& a^2-\varepsilon_2
	\end{pmatrix}
   ,-\varepsilon_3,\ldots,-\varepsilon_n
\right)
\end{equation}
	Consequently, due to
 $\varepsilon_1+\varepsilon_2=0$ and $\varepsilon_1\varepsilon_2=-1$,
the eigenvalues of $[h_{ij}]$, which are the roots of
 $\det[h_{ij}-\lambda\delta_{ij}]=0$,
are $\lambda_i=-\varepsilon_i$ for $i\geq3$ and
 $\lambda_\pm = a^2 \pm \sqrt{1+a^4} \gtrless 0$.
So, in this case the signature of $h$ is
 $(q-1,p-1)+(1,1)=(q,p)$ as that of $g$ is $(p,q)$.
	\end{Proof}

	\begin{Cor}	\label{Cor3.1}
Let $g$ be a Riemannian metric of signature $(p,q)$ on~$U\subseteq M$ and~$t$
be a vector field on $U$. Assume $t$ can be chosen such that $g(t,t)$ is less
than, or greater than, or equal to one on the whole set $U$. Then on $U$ the
tensor field $h$ given by~\eref{3.1} is:

\textup{\textbf{\hphantom{ii}(i)}}
a Riemannian metric with signature $(q,p)$ for $g(t,t)<1$.

\textup{\textbf{\hphantom{i}(ii)}}
a Riemannian metric with  signature $(q+1,p-1)$ for $g(t,t)>1$.

\textup{\textbf{(iii)}}
a (parabolic) semi-Riemannian metric with signature $(q,p-1)$
and defect $1$ for $g(t,t)=1$, \ie in this case $h$ has $q$ positive,
$(p-1)$ negative, and 1 vanishing eigenvalue.
	\end{Cor}

	\begin{Proof}
This result is a version of proposition~\ref{Prop3.1} corresponding to the
choice of $t$ such that one of the sets $\widetilde{U}_{<}^{+}$,
$\widetilde{U}_{>}^{+}$, and $\widetilde{U}_{=}^{+}$ is equal to $U$.
	\end{Proof}

	Evidently, the main results formulated in sections~\ref{Sect2}
and~\ref{applicability} are special cases, corresponding to $(p,q)=(n,0)$, of
the just proved proposition~\ref{Prop3.1} and corollary~\ref{Cor3.1}.

It is clear that if $g$ is a Riemannian metric on $U$, then, choosing
arbitrary some vector field $t$ on $U$ with $g(t,t)>1$, the map~\eref{3.1}
yields (infinitely) many (semi\ndash)Riemannian metrics on $U$ whose
signature (and, possibly, defect) depends on the norm $g(t,t)$ on $U$.
Generally different $t$ generate different metrics $\Tilde{g}^+$  from one
and the same initial metric $g$.

	At this point a natural question arises: Does a vector field $t$ with
the properties described in corollary~\ref{Cor3.1} exist? Here, when $U=M$,
the situation is completely the same as described in
Sect.~\ref{applicability} for the Euclidean case ($g=e$). If on $t$ are not
imposed some additional, \eg smoothness, conditions, a vector field $t$ on $U$
with $g(t,t)|_U\gtreqless1$ can always be constructed for every $U\subseteq M$.
In fact, let $t_0$ be any (generally discontinuous) non\ndash vanishing on
$U$  vector field. By rescaling locally the components of $t_0$ we can obtain
from it a non\ndash vanishing vector field $t_{0}^{\prime}$ such that
 $g(t_{0}^{\prime},t_{0}^{\prime})|_U\not=0$
and
 $\sign(g(t_{0}^{\prime},t_{0}^{\prime})|_U)=\varepsilon=\const$.
Defining
\(
t := \sqrt{a}t_{0}^{\prime}
	\lvert g(t_{0}^{\prime},t_{0}^{\prime})\rvert^{-1/2}
\)
for $a\in\mathbb{R}$, $a\ge0$, we get $g(t,t)=\varepsilon a$.
Consequently, by an appropriate choice of $\varepsilon$ and $a$, we can
realize $t$ with $g(t,t)|_U\gtreqless 1$. Since $g$ is by definition
nondegenerate (the kernel of $g$ consists of the null vector field on $U$),
the relation $g(t,t)|_U\ge 1$ implies $t$ to be non\ndash vanishing vector
field. Obviously, this conclusion does not concern the case of $t$ with
$g(t,t)|_U<1$ when $t$ can vanish somewhere or everywhere on $U$.

	As we know from Sect~\ref{applicability}, the situation for $t$ with
$g(t,t)|_U\ge 1$ is completely different when $U=M$ and $C^m$, $m\ge0$,
metrics and vector fields are considered: Generally such vector field does
not exist globally, \ie on the whole manifold $M$. This existence depends on
the topological properties of $M$ and has to be investigated separately in
any particular case.

	Now consider the class of (resp.\ smooth) Lorentzian metrics on $M$,
\ie those $g$ for which $(p,q)=(1,n-1)$. For them, according to
corollary~\ref{Cor3.1}, the metric $h$ is of signature
 $(n-1,1)$ for $g(t,t)<1$ and
 $(n,0)$ for $g(t,t)>1$ (resp.\ if such $t$ exists on $M$),
\ie in the former case $g$ and $h$ are isomorphic and in the latter one $h$
is Euclidean metric. Thus, if for every $g$ we choose some vector field $t_g$
with $g(t_g,t_g)>1$, then the whole class of (resp.\ smooth) Lorentzian
metrics is mapped into the class of (resp.\ smooth) Euclidean ones by the
mapping $g\mapsto h$ given by~\eref{3.1} for $t=t_g$ (resp.\ if such
smooth $t_g$ exists on $M$). Evidently, different vector fields $t_g$ realize
different such maps.

	Now some natural questions are in order. Let $G^U$  (resp.\
$G_{p,q}^{U}$) be the set of all Riemannian metrics (resp.\ of signature
$(p,q)$) on $U\subseteq M$. If $t$ is a fixed vector field on $U$, then what
is the character of the map $\varphi_{U}^{t}\colon G^U\to G^U$ given
by~\eref{3.1}? For instance, can it be surjective, injective, or bijective?
Can any two Riemannian metrics (with `corresponding' signatures) be mapped
into each other by $\varphi_{U}^{t}$ for a suitable $t$? Etc.

	\begin{Prop}	\label{Prop3.2}
Let $g\in G_{p,q}^{U}$, $t$ be arbitrarily fixed vector field on
$U$, and $\varphi_{U}^{t}\colon G^U\to G^U$ be given via~\eref{3.1}. Then:

\textbf{(i)}
The map $\left.\varphi_{U}^{t}\right|_{t=0}$ is bijection.

\textbf{(ii)}
If $t\not=0$, the map $\varphi_{U}^{t}$ is injection on the sets
 $\{ g : g\in G_{p,q}^{U},\quad g(t,t) = 1/2 \}$ and
 $\{ g : g\in G_{p,q}^{U},\quad g(t,t) = 0 \}$.

\textbf{(iii)}
If $g(t,t)\not=0,\frac{1}{2}$,
then there exists a metric $g^\prime\in G_{p,q}^{U}$ such
that  $\varphi_{U}^{t}(g)=\varphi_{U}^{t}(g^\prime)$ and $g\not=g^\prime$.
Besides, this $g^\prime$ is unique iff $n:=\dim M=1$ or $n\geq2$ and
$g(t,t)\not=1$, \ie $\varphi_{U}^{t}$ is two\ndash{to}\ndash{one} mapping on
the sets
 $\{ g : g\in G_{p,q}^{U},\quad g(t,t) \not= 0, \frac{1}{2} \}$
for $n=1$ and
$\{ g : g\in G_{p,q}^{U},\quad g(t,t) \not= 0, \frac{1}{2}, 1 \}$
for $n\geq2$.
For $n\geq2$ and $g(t,t)=1$ this metric $g^\prime$ depends on $(n-1)$
non\ndash zero real parameters.
	\end{Prop}

	\begin{Proof}%>>>>>>>>>>>>>>>>>>>>>>>>>>>>>>>>>>>>>>>>>>>>>>>>>>>>>>
\textbf{Case (i)}.
For $t=0$ we have $\varphi_{U}^{t}(g)=-g$, so $\varphi_{U}^{0}$ is reversing
of the the metric~\cite[p.~92]{O'Neill} and hence it is bijective.

\textbf{Cases (ii) for $g(t,t)=1/2$ and (iii)}.
Consider the equation $\varphi_{U}^{t}(g)=\varphi_{U}^{t}(g^\prime)$,
$t\not=0$ with respect to $g^\prime$. In the special basis (at some $x\in U$
and with respect to $g$) defined in the `non\ndash isotropic' case of the
proof of proposition~\ref{Prop3.1}, this equation reads
	\begin{equation}	\label{3.2}
\alpha^2( \delta_{i1}\delta_{j1} - g_{i1}^{\prime}g_{j1}^{\prime} )
 = \varepsilon_i\delta_{ij} - g_{ij}^{\prime},
\qquad
\varepsilon_{i} = \pm 1,	\quad \alpha\not=0 .
	\end{equation}
In the same basis we also have $g_{11}=\varepsilon_1$ and
 $g(t,t)=\varepsilon_1\alpha^2$.

	At first we consider the one-dimensional case, $n:=\dim M=1 $.
Now, multiplying~\eref{3.2} with $\varepsilon_1$, we get
\(
g(t,t) (1-\varepsilon_1g_{11}^{\prime}) (1+\varepsilon_1g_{11}^{\prime})
 = 1 - \varepsilon_1g_{11}^{\prime}.
\)
So, the last equation has two solutions with respect to $g_{11}^{\prime}$:
 $g_{(1)\, 11}^{\prime}=\varepsilon_1=g_{11}$ and
\(
g_{(2)\, 11}^{\prime}
 = \big( \frac{1}{g(t,t)} -1 \bigr) \varepsilon_1
 = \big( \frac{1}{g(t,t)} -1 \bigr) g_{11}
\)
By virtue of $n=1$, this means that only two metrics $g^\prime$
satisfy~\eref{3.2}:
 $g_{(1)}^{\prime}=g$ and $g_{(2)}^{\prime}=\big(\frac{1}{g(t,t)}-1\bigr)g$.

	Evidently $g_{(2)}^{\prime}\not=g$ iff $g(t,t)\not=1/2$ which
completes the proof for $n=1$ of the cases (ii) for $g(t,t)=1/2$ and (iii).

	Let now $n\geq2$.

	For $i=j=1$ equation~\eref{3.2} yields
	\begin{equation}	\label{3.3}
\alpha^2\bigl( 1-(g_{11}^{\prime})^2 \bigr)
 = \varepsilon_1 - g_{11}^{\prime}
	\end{equation}
with solutions
	\begin{equation}	\label{3.3'}	\tag{\ref{3.3}$^\prime$}
g_{(1)\, 11}^{\prime}=\varepsilon_1=g_{11}
\qquad\text{and}\qquad
g_{(2)\, 11}^{\prime}
 = \left( \frac{1}{g(t,t)} -1 \right) g_{11} .
	\end{equation}

	For $i,j\geq2$ equation~\eref{3.2} reduces to
	\begin{equation}	\label{3.4}
-\alpha^2 g_{i1}^{\prime}g_{j1}^{\prime}
 = \varepsilon_i\delta_{ij} - g_{ij}^{\prime},
\qquad i,j\geq2.
	\end{equation}

	At last, for $i\not=1$, $j=1$ or $i=1$, $j\not=1$ equation~\eref{3.2}
gives
$\alpha^2 g_{i1}^{\prime}g_{11}^{\prime} = g_{i1}^{\prime}$,
the solutions of which are
	\begin{subequations}
\begin{equation}	\label{3.5a}
g_{i1}^{\prime} = 0 \quad(=g_{i1}), \qquad i\geq2
\end{equation}
and
\begin{equation}	\label{3.5b}
g_{11}^{\prime} = \frac{1}{\alpha^2},\quad g_{i1}^{\prime}\not=0
\quad\text{for $i\geq2$} .
\end{equation}
	\end{subequations}

	Consider the case~\eref{3.5a}. Now~\eref{3.4} reads
 $g_{ij}^{\prime}=\varepsilon_i\delta_{ij}=g_{ij}$ for $i,j\geq2$. So
(see~\eref{3.5a}) $g_{ij}^{\prime}=g_{ij}$ for $(i,j)\not=(1,1)$.
Combining this with~\eref{3.3'}, we find the following two solutions for
$g^\prime$:
\[
g_{(1)}^{\prime} =g
\quad\text{and}\quad
g_{(2)\, ij}^{\prime} =
	\begin{cases}
	g_{ij}&					\text{for $(i,j)\not=(1,1)$}\\
	\big(\frac{1}{g(t,t)}-1\bigr)g_{11}&	\text{for $(i,j)=(1,1)$}
	\end{cases}
\quad .
\]
Evidently $g_{(2)}^{\prime}\not=g$ iff $g(t,t)\not=1/2$ which proves the
cases under consideration for $g(t,t)=1/2$ and $n\geq2$.

	The proof of the rest of case (iii) is a consequence of
case~\eref{3.5b}. Substituting~\eref{3.5b} into~\eref{3.3}, we get
$\alpha^2=\varepsilon_1$ and, consequently, as $\alpha\in\mathbb{R}$, this is
possible if and only if $g(t,t)=1=\varepsilon_{1}^{2}=\alpha^2$. If this is
the case, equation~\eref{3.4} yields
\(
g_{ij}^{\prime}
 = \varepsilon_i\delta_{ij} + g_{i1}^{\prime} g_{j1}^{\prime},
\)
 $i,j\geq2$. Combining the last results with~\eref{3.5b}, we obtain for
$g(t,t)=1$ the solution
\[
g_{(3)\, ij}^{\prime}
 =
	\begin{cases}
g_{11}=1		&\text{for $i,j=1$}		\\
g_{ij} +a_ia_j		&\text{for $i,j\geq2$}		\\
a_i			&\text{for $i\geq2$, $j=1$}	\\
a_j			&\text{for $i=1$, $j\geq2$}
	\end{cases}
\]
where $a_i$, $i\geq2$ are arbitrary non\ndash{zero} real numbers. As
$a_i\not=0$ (see~\eref{3.5b}), we have
$0\not=g_{(3)\, i1}^{\prime}\not=g_{i1}=0$ for $i\geq2$.  Therefore, if
$\dim M\geq2$ and $g(t,t)=1$, the initial equation admits an $(n-1)$
parameter family of solutions $g^\prime\not=g$

	At last, we have to consider the
\textbf{case (ii) for $g(t,t)=0$ with $t\not=0$}.
This is possible only for $n\geq2$ (see the last part of the proof of
proposition~\ref{Prop3.1}).

	In the special basis $\{E_i\}$ introduced in the `general isotropic'
case of the proof of proposition~\ref{Prop3.1}, the equation
 $\varphi_{U}^{t}(g)=\varphi_{U}^{t}(g^\prime)$, after some algebra, takes
the form
	\begin{equation}	\label{3.7}
a^2
\bigl[
\varepsilon_i\varepsilon_j(\delta_{i1}+\delta_{i2}) (\delta_{j1}+\delta_{j2})
-
(g_{i1}^{\prime}+g_{i2}^{\prime})(g_{j1}^{\prime}+g_{j2}^{\prime})
\bigr]
 =
\varepsilon_i\delta_{ij} - g_{ij}^{\prime}.
	\end{equation}

	For $i,j=1,2$ this equation reduces to
\[
a^2\bigl[
 1 + (g_{11}^{\prime} + g_{12}^{\prime}) (g_{22}^{\prime} + g_{12}^{\prime})
\bigr]
 =
g_{12}^{\prime},
\qquad
a^2\bigl[ 1 - (g_{ii}^{\prime} + g_{12}^{\prime})^2 \bigr]
=
\varepsilon_i - g_{ii}^{\prime}
% , \quad  i=1,2
\]
as $\varepsilon_1+\varepsilon_2=0$ and $\varepsilon_{i}^{2}=1$.
These equations form a closed system for $g_{ij}^{\prime}$ with $i,j=1,2$.
It has a unique solution. Actually, putting for brevity
 $x_i:=g_{i1}^{\prime}+g_{i2}^{\prime}$, $i=1,2$ and $z:=g_{12}^{\prime}$, we
see that the equations mentioned can be written as
\(
z=a^2(1+x_1x_2)\text{ and }
a^2x_{i}^{2}-x_i+\varepsilon_i-a^2+z=0,
\ i=1,2
\)
and hence
\(
a^2x_{i}^{2}-x_i+\varepsilon_i-a^2x_1x_2=0,
\ i=1,2 .
\)
Forming the sum and difference of the last two equations, we, by virtue
of $\varepsilon_1+\varepsilon_2=0$, find
\(
a^2(x_1+x_2)^2 - (x_1+x_2) = 0
\text{ and }
a^2(x_1+x_2)(x_1-x_2) - (x_1-x_2) + \varepsilon_1-\varepsilon_2 = 0
\)
respectively.
	The first of these equations has two solutions,  $(x_1+x_2)=0$ and
$(x_1+x_2)=1/a^2$, but only the first of them, $(x_1+x_2)=0$, is compatible
with the second of the above equations due to
$\varepsilon_1+\varepsilon_2=0$, $\varepsilon_{1,2}=\pm1$. In this way we see
that $(x_1+x_2)=0$ and $(x_1-x_2)=\varepsilon_1-\varepsilon_2$. Therefore we
obtain $x_1=\varepsilon_1$, $x_2=\varepsilon_2$, and
$z=a^2(1+\varepsilon_1\varepsilon_2)=0$. Returning to the initial variables,
we, finally, get
\[
g_{ij}^{\prime}=g_{ij}=\varepsilon_i\delta_{ij}
\quad\text{for $i,j=1,2$} .
\]

	Consider now~\eref{3.7} for $i=1,2$  and $j\geq3$:
\(
a^2 (g_{i1}^{\prime}+g_{i2}^{\prime})(g_{j1}^{\prime}+g_{j2}^{\prime})
 =
g_{ij}^{\prime}.
\)
By virtue of the last result it reduces to
\(
a^2\varepsilon_i (g_{j1}^{\prime}+g_{j2}^{\prime}) = g_{ij}^{\prime},
\)
 $i=1,2$, $j\geq3$.
Summing the equations corresponding to $i=1,2$ and using
$\varepsilon_1+\varepsilon_2=0$, we find $g_{1j}^{\prime}+g_{2j}^{\prime}=0$
for $j\geq3$, which, when inserted into the initial equations, leads to
\[
g_{ij}^{\prime} = g_{ij} =0,
\qquad i=1,2,\quad j\geq3.
\]

	At last, for $i,j\geq3$ equation~\eref{3.7} gives
\(
- a^2
(g_{i1}^{\prime}+g_{i2}^{\prime})(g_{j1}^{\prime}+g_{j2}^{\prime})
 =
\varepsilon_i\delta_{ij} - g_{ij}^{\prime}.
\)
Substituting here the last result, we find
\[
g_{ij}^{\prime}=g_{ij}=\varepsilon_i\delta_{ij}
\qquad\text{for $i,j\geq3$}.
\]

	Combining the last three results, we, finally, obtain $g^\prime=g$,
\ie the only solution of~\eref{3.7} is $g^\prime=g$.
Hence, in the case considered the map $\varphi_{U}^{t}$ is an injection.
	\end{Proof}%<<<<<<<<<<<<<<<<<<<<<<<<<<<<<<<<<<<<<<<<<<<<<<<<<<<<<<<

	\begin{Rem}	\label{Rem3.0}
From the proof of proposition~\ref{Prop3.2} follows that in the first subcase
of case~(iii) of proposition~\ref{Prop3.2}, when $\varphi_{U}^{t}$ is~2:1
map, is fulfilled
 $g^\prime(t,t)=1-g(t,t)$ (with $g^\prime=g_{(2)}^\prime$)
while in the second subcase of case~(iii) is valid
 $g^\prime(t,t)=g(t,t)$ (with $g^\prime=g_{(3)}^\prime$).
	\end{Rem}

	\begin{Prop}	\label{Prop3.3}
Let the  vector field $t$ be arbitrarily fixed on $U$ and the map
$\varphi_{U}^{t}\colon G^U\to G^U$ be given by~\eref{3.1}.
Then $(\varphi_{U}^{t}\circ\varphi_{U}^{t})(g)=g$
iff $g\in G^U$ is such that $g(t,t)=0,2$.
	\end{Prop}

	\begin{Rem}	\label{Rem3.1}
Note, due to~\eref{3.1}, we have
	\begin{equation}	\label{3.0+2}
\bigl(\varphi_{U}^{t}(g)\bigr) (t,t) = g(t,t)
\qquad \text{iff $g(t,t)=0,2$}.
	\end{equation}
	\end{Rem}

	\begin{Proof}
	Consider the case $g(t,t)\not=0$. In a basis in which~\eref{3.h}
holds, equation~\eref{3.1} gives ($h:=\varphi_{U}^{t}(g)$)
\[
\bigl[ (\varphi_{U}^{t}(h))_{ij}\bigr]
 =
\diag\big( \varepsilon_1[g(t,t)-1][g(t,t)(g(t,t)-1)-1],
				\varepsilon_2,\dots,\varepsilon_n \big)
\]
because of
 $t^i=\alpha\delta^{i1}$, $g(t,t)=\varepsilon_1\alpha^2$,
 $t_{i}^{g}:=g_{ij}t^j=\varepsilon_1\alpha\delta_{i1}$, and
 $t_{i}^{h}:=h_{ij}t^j=\varepsilon_1\alpha(g(t,t)-1)\delta_{i1}$.
A simple computation shows that the last matrix is equal to
 $g_{ij}=\diag(\varepsilon_1,\dots,\varepsilon_n)$
iff $g(t,t)=2$ as we supposed $g(t,t)\not=0$.

If $t=0$, then
$(\varphi_{U}^{0}\circ\varphi_{U}^{0})(g)=+\varphi_{U}^{0}(-g)=g$ for every
$g\in G^U$ due to~\eref{3.1}.

	At last, let $t\not=0$ and $g(t,t)=0$. In a basis in
which~\eref{3.h-isotropic} holds, from~\eref{3.1}, we obtain
 $[(\varphi_{U}^{t}(h))_{ij}] = [g_{ij}]$
due to
 $t^i=a(\delta^{i1}+\delta^{i2})$,
 $t_{i}^{g}:=g_{ij}t^j=a\varepsilon_i(\delta_{i1}+\delta_{i2})$, and
\(
t_{i}^{h}:=h_{ij}t^j
 = -a(\varepsilon_1\delta_{i1} + \varepsilon_2\delta_{i2})$.
	\end{Proof}

	From the just proved result immidiately follows (see
also~\cite[p.~14, proposition~6.9]{Dugundji-topology})

	\begin{Cor}	\label{Cor3.2}
The map $\varphi_{U}^{t}$ for given $t$ is bijective on the sets
 $G_{t;2}^{U}:=\{g: g\in G^U,\ g(t,t)=2\}\subset G^U$ and
 $G_{t;0}^{U}:=\{g: g\in G^U,\ g(t,t)=0\}\subset G^U$.
	\end{Cor}

	\begin{Rem}	\label{Rem3.2}
The bijectiveness of $\varphi_{U}^{t}$ on $G_{t;2}^{U}$ does not contradict to
proposition~\ref{Prop3.2}, case~(ii). Actually, if $g\in G_{t;2}^{U}$,
 $g^\prime\in G^{U}$, $g\not=g^\prime$, and
$\varphi_{U}^{t}(g^\prime)=\varphi_{U}^{t}(g)$,
then (see remark~\ref{Rem3.0})
 $g^\prime(t,t)=1-g(t,t)=-1\not=2$,
\ie $g^\prime$ is not in $G_{t;2}^{U}$.
	\end{Rem}

	Ending this section, we have  to note that if the Riemannian metrics
$g$ and $h$ are given, then generally there does not exists a vector field
$t$ connecting them through
$h=g(\cdot,t)\otimes g(\cdot,t) - g$.
There are two reasons for this. On one hand, by propositon~\ref{Prop3.1} for
this the metrics  $g$ and $h$ must be of `corresponding' signature, \viz
 $(p,q)$ and $(q+1,p-1)$ or $(p,q)$ and $(q,p)$ respectively. On the other
hand, in local coordinates the mentioned connection between $g$ and $h$
reduces to a system of $n(n+1)/2$ equations for the  $n$ components of $t$
and, consequently, it has solution(s) only in some exceptional cases. It is
clear that even for Euclidean metric $g$ and Lorentzian metric $h$ such $t$
exixts only as an exception, not in the general case.

\section {Consequences}
\label{Sect4}
\setcounter {equation} {0}

	Since the case (ii) of corollary~\ref{Cor3.1} is most important
in connection with possible applications, we investigate below some
consequences of it.

	Let $G_{p,q}^{U}$ be the set of all Riemannian metrics with signature
$(p,q)$ on $U\subseteq M$.

	\begin{Cor}	\label{Cor4.1}
Let $p\geq1$, for every $g\in G_{p,q}^{U}$ be chosen a vector field $t_g$ on
$U$ such that $g(t_g,t_g)=2$, and $T:=\{t_g : g\in G_{p,q}^{U}\}$. Then the
map $\varphi_{p,q}^{T}\colon G_{p,q}^{U}\to G_{q+1,p-1}^{U}$ given via
	\begin{equation}	\label{4.1}
g\mapsto\Tilde{g}^+ := g(\cdot,t_g)\otimes g(\cdot,t_g) - g
 	\end{equation}
is bijective, \ie one-to-one onto map.
	\end{Cor}

	\begin{Proof}
At first we note that $t_g$ with $g(t_g,t_g)=2$ always exists for every $g$
because of $p\geq1$. (E.g.\ one can set $t_g=\sqrt{2}t_0/\sqrt{g(t_0,t_0)}$,
where in a basis in which $g_{ij}=\varepsilon_i\delta_{ij}$,
$\varepsilon_i=\pm1$, and $\varepsilon_k=+1$ for some fixed
$k\in\{1,\ldots,n\}$ the components of $t_0$ are
$t_{0}^{i}=\alpha\delta^{ik}$, $\alpha\in\mathbb{R}\backslash\{0\}$; so then
$g(t_0,t_0)=\alpha^2>0$.) Now, from proposition~\ref{Prop3.3} and
corollary~\ref{3.1}, case~(ii), we deduce
	\begin{equation}	\label{4.5}
\varphi_{q+1,p-1}^{T}\circ \varphi_{p,q}^{T} = \id_{G_{p,q}^{U}}, \quad
\varphi_{p,q}^{T}\circ \varphi_{q+1,p-1}^{T} = \id_{G_{q+1,p-1}^{U}} .
	\end{equation}
from where, by virtue of~\cite[p.~14, proposition~6.9]{Dugundji-topology},
the result formulated follows.
	\end{Proof}

	In this way, by an explicit construction, we proved the existence of
a bijective correspondence between the classes of Riemannian metrics
with signature $(p,q)$ and $(q+1,p-1)$ on any differentiable manifold
admitting such metrics (and vector fields with corresponding properties -
see Sect.~\ref{applicability}). It has to be emphersized on the explicite
dependence of this mapping on the choice of the vectors $t_g$ utilized in its
construction. In particular, which is essential for the physics, there is a
bijective correspondence between the sets of Euclidean and Lorentzian
metrics as they have signatures $(n,0)\text{ and } (1,n-1)$ respectively.%
\footnote{%
In the 4-dimensional case, a special type of relation between Euclidean and
Lorentzian metrics is established in~\cite{Barbero96} via the Einstein
equations.%
}

	From here an important result follows.
	Since every paracompact finite\ndash dimensional differentiable
manifold admits Euclidean
metrics~\cite[chapter~IV, \S~1, chapter~I, example~5.7]{K&N-1},
\cite[p.~280]{Bruhat},
\emph{%
on any such manifold admitting a vector field with an Euclidean
norm greater then one exist Lorentzian metrics%
}
as they are in bijective correspondence with the Euclidean ones.%
\footnote{%
See~\cite{Markus} and~\cite[p.149, proposition~37]{O'Neill} for more general
results on the existens of Lorentzian metrics.%

}
	The opposite statement is also true: if on $M$ exist Lorentzian, $h$,
and Euclidean, $e$, metrics, then there is a vector field $t$ with
$e(t,t)>1$.%
\footnote{
Generally $h,\ e$, and $t$ are not connected via~\eref{2.1}
(see the remark at the end of Sect~\ref{Sect4}).%
}
In fact, since $h$ is Lorentzian, there is exactly one positive
eigenvalue $\lambda_+$, $\lambda_+>0$, for which the equation
 $h_{ij}t_{+}^{j} = \lambda_+e_{ij}t_{+}^{j}$
has a non-zero solution $t_+$ defined up to a non-zero constant multiplier.
Choosing this constant such that $h(t_+,t_+)>\lambda_+$, we find
$e(t_+,t_+)>1$. Let us recall (see Sect.~\ref{applicability}) that the
existence of $t$ with $e(t,t)>1$ is equivalent to the one of a non-vanishing
vector field on $M$. So, if, as usual, we admit $e,\ h$, and $t$ to be of
class $C^m$, $m\geq0$, than such a vector field may not exist on the whole
$M$. If this happens to be the case, the above, as well as the following,
considerations have to be restricted on the neighborhood(s) admitting
non-vanishing vector field of class $C^m$.

	Since~\eref{3.1} is insensitive to the change $t\mapsto-t$, we are
practically dealing with the field $(t,-t)$ of linear elements,
\ie\cite[sect.~2.6]{Hawking&Ellis} a field of pairs of vector fields with
opposite directions, not with the vector field $t$ itself. If $(X,-X)$ is a
field of linear elements on $M$, then for any $\lambda\in\mathbb{R}$,
$\lambda>1$ the vector fields $t_\pm:=\pm\sqrt{\lambda/e(X,X)}X$ have
Euclidean norm $e(t_\pm,t_\pm)=\lambda>1$. Conversely, if $t$ is a vector
field with $e(t,t)>1$, then $(t,-t)$ is a field of linear elements on $M$.
Combining the just-obtained results, we infer that on $M$ exist Lorentzian
metrics iff on it exists a field of linear elements. This is a known result
that can be found, e.g, in~\cite[sect.~2.6]{Hawking&Ellis}.

	Let $e$ and $h$ be respectively Euclidean and Lorentzian metrics
connected by~\eref{3.1} for some $t$ with $e(t,t)>1$. Now we shall prove that
for a suitable choice of $t$ the set $V$ of vector fields on $M$ can be split
into a direct sum $V=V^+\oplus V^-$ in which $V^+$ is orthogonal to $V^-$
with respect to both $e$ and $h$, and
$\left.h\right|_{V^\pm}=\pm\left.e\right|_{V^\pm}$.
In fact, defining
 $V^+:=\{ t^+ : t^+=\lambda t,\ \lambda\in\mathbb{R}\backslash\{0\} \}$ and
 $V^-:=\{ t^- : e(t^-,t)=0 \}$,
we see that for $s^\pm,t^\pm\in V^\pm$ is fulfilled
 $e(t^-,t^+) = h(t^-,t^+) \equiv 0$,
 $h(s^-,t^-) \equiv - e(s^-,t^-)$ and
 $h(s^+,t^+) = (e(t,t)-1) e(s^+,t^+)$.
The choice of $t$ with $e(t,t)=2$ completes the proof.
	In this way we have obtained an evident special case, concerning
Lorentzian metrics, of~\cite[p.~434, proposition~VII]{Greub&et.al.-2}.
As a consequence of the last proof, as well as of~\eref{3.1}, we see that any
set of vector fields in $V^-$ which are mutually orthogonal (or orthonormal)
with respect to $e$ is such also with respect to $h$ for any $t$ with
$e(t,t)>1$ (a good choice is $e(t,t)=2$ - see~\eref{3.0+2}.
Sets of this kind are often used in physics~\cite{Hawking&Ellis}. Evidently,
if we add to such a set the victor field $t$, the mutual orthogonality of the
vector fields of the new set will be preserved.

	Another significant corollary from the proved equivalence between
Lo\-rent\-zian and Euclidean metrics is that any physical theory formulated
in terms of (real) Lorentzian metric(s), e.g.\ the special theory of
relativity or relativistic quantum mechanics, can equivalently be
(re)formulated in terms of (real) Euclidean metric(s).%
\footnote{%
For instance, in the four-dimensional case, $n=4$, in an appropriately chosen
local coordinates in which the Euclidean and Lorentzian metrics are
represented respectively by the unit matrix and Minkowski metric tensor, \ie
$[e_{ij}]=\diag(1,1,1,1)$ and $[\eta_{ij}]=\diag(1,-1,-1,-1)$, the connection
between them is $\eta_{ij}=t_it_j-e_{ij},\ t_i:=e_{ij}t^j=t^i$ with
$t^i=\sqrt{2}\delta^{i1}$ in the coordinates used.%
}
The price one pays for this is the introduction of an additional vector field
$t$ whose physical meaning is not a subject of this paper.

	So, in some sense, the deviation of a Lorentzian metric $g$ from
an Euclidean one $e$ can be described by an appropriate choice of
certain vector field $t$, all connected by~\eref{2.2} under the condition
$e(t,t)>1$.  In~\cite{Pestov96} this vector field is interpreted as a field
of the time, the so called temporal field. In~\cite{Pestov96} on $t$ is
imposed the normalization condition $e(t,t)=1$
(see~\cite[equation~(3)]{Pestov96}) which, as we proved in this paper,
contradicts to the Riemannian character of the metrics considered.
Consequently, this condition has to be dropped and replace with $e(t,t)>1$.
The physical interpretation of $t$ as a temporal field will be studied
elsewhere.

	We also have to note that the statement in~\cite[p.~13]{Pestov96} that
the determinants of corresponding via~\eref{2.1} Euclidean and Lorentzian
metrics differ only by sign is generally wrong. In fact, in a special basis
in which~\eref{2.3} holds, we have
$\det[g_{ij}]=(-1)^{n+1}(e(t,t)-1)$
which in an arbitrary basis reads
$\det[g_{ij}]=(-1)^{n+1}(e(t,t)-1)\det[e_{ij}]$.
Therefore  $\det[g_{ij}] +\det[e_{ij}] = 0$
%According to~\eref{2.3}, this
is true only in two special cases, viz.\
if $n=2k\text{ and }e(t,t)=2$ or
if $n=2k+1\text{ and }e(t,t)=0$, $k=0,1,\ldots$
Moreover, by corollary~\ref{Cor3.1}, the second case cannot be
realized if $e$ is Euclidean and $g$ Lorentzian. Thus the mentioned statement
is valid only on even-dimensional manifolds and vector fields $t$ with
norm~2.

\section {Conclusion}
\label{conclusion}
\setcounter{equation} {0}

	The main results of the previous considerations are expressed by
propositions~\ref{Euclidean} and~\ref{Prop3.1}--\ref{Prop3.3}. As we saw in
Sect.~\ref{Sect4}, their consequence is the existence of bijective mapping
between metrics of signatures $(p,q)$ and $(q+1,p-1)$, in particular between
Euclidean and Lorentzian metrics. Consequently, a physical theory formulated
in terms of real Lorentzian metric(s) on a manifold admits, possibly
locally, equivalent (re)formulation in terms of real Euclidean metric(s) on
the same manifold. Another corollary of these propositions is that on a
manifold exist metrics of signature $(q+1,p-1)$ if it admits a metric $g$ of
signature $(p,q)$ and a vector field $t$ with $g(t,t)>1$. When applied to
Lorentzian and Euclidean metrics, the last assertion reproduces a known
result~\cite[sect.~2.6]{Hawking&Ellis}. Vector fields $t$ with $g(t,t)>1$
exist on $M$ iff it admits a non-vanishing vector field over the whole
manifold $M$. If we do not impose additional conditions on the last field it
always exists. But if we require it to be of class $C^m$ with $m\geq0$ its
existence is connected with the topological properties of $M$ and has to be
explored in any particular case. Generally non-vanishing $C^m$ vector fields
exist locally, but globally this may not be the case.

	Analogous consequences can be made from case (i) of
corollary~\ref{Cor3.1}.

	\begin{Cor}	\label{CorConcl.0}
Let for every $g\in G_{p,q}^{U}$ be chosen a vector field $t_g$ on $U$ such
that $g(t_g,t_g)=0$  and $T:=\{t_g: g\in G_{p,q}^{U}\}$. Then the map
 $\psi_{p,q}^{T}\colon G_{p,q}^{U}\to G_{q,p}^{U}$ defined by~\eref{4.1} is
a bijection.
	\end{Cor}

	\begin{Proof}
At the beginning we notice that one can always put $t_g=0$ for every
$g\in G_{p,q}^{U}$ but if $p,q\geq1$, then for any $g\in G_{p,q}^{U}$ exists
$t_g\not=0$ with $g(t_g,t_g)\not=0$. (In a basis in which
$g_{ij}=\varepsilon_i\delta_{ij}$, $\varepsilon_i=\pm1$ and
$\varepsilon_k+\varepsilon_l=0$ for some fixed $k,l\in\{1,\ldots,n\}$ we can
set $t_g^i=\alpha(\delta^{ik}+\delta^{il})$,
$\alpha\in\mathbb{R}\backslash\{0\}$.) From proposition~\ref{Prop3.3} and
corollary~\ref{Cor3.1}, case~(i), we infer

	\begin{equation}	\label{Concl.0}
\psi_{p,q}^{T}\circ \psi_{q,p}^{T} = \id_{G_{q,p}^{U}},
\qquad
\psi_{q,p}^{T}\circ \psi_{p,q}^{T} = \id_{G_{p,q}^{U}}
	\end{equation}
which concludes the proof.
	\end{Proof}

	Hence there is bijective correspondence between metrics of signature
$(p,q)$ and $(q,p)$. Etc. It is important to be noted that the case of a
vector field $t$ with $g(t,t)<1$ significantly differs from the one of $t$
with $g(t,t)\geq1$ when some smoothness conditions are imposed: $C^m$,
$m\geq0$ vector field $t$ with $g(t,t)<1$ exists over any subset
$U\subseteq M$, in particular over the whole manifold $M$. In fact, a
trivial example of this kind is the null vector field over $U\subseteq M$.

	Let us fix some bijective maps
 $\varphi_{p,q}\colon G_{p,q}^{U}\to G_{q+1,p-1}^{U}$
and
 $\psi_{p,q}\colon G_{p,q}^{U}\to G_{q,p}^{U}$
given via~\eref{4.1} for $t_g$ with
 $g(t_g,t_g)=2$ and $g(t_g,t_g)=0$ respectively.
Here $G_{p,q}^{U}$ is the set of Riemannian metrics on $U$ with signature
$(p,q)$. (Let us recall that in the `smooth' case we can not put $U=M$ as,
generally, then $\varphi_{p,q}$ may not exist.) Then the map
\(
\chi_{p,q} := \psi_{q+1,p-1}\circ\varphi_{p,q}
	\colon G_{p,q}^{U} \to G_{p-1,q+1}^{U}
\)
is bijective for any $p,q\in\mathbb{N}\cup\{0\}$ such that $p+q=n:=\dim M$.
Hence
\[\begin{CD}
G_{n,0}^{U}	@>\chi_{n,0}>>	 G_{n-1,1}^{U}	@>\chi_{n-1,1}>>
G_{n-2,2}^{U}	@>\chi_{n-2,2}>> \cdots		@>\chi_{2,n-2}>>
G_{1,n-1}^{U}	@>\chi_{1,n-1}>> G_{0,n}^{U}
\end{CD}\]
is a sequence of bijective maps. In short, this means that there is an
bijective real correspondence (given explicitly via compositions of maps
like~\eref{3.1}) between Riemannian metrics of arbitrary signature.
Therefore, starting from the class of Euclidean metrics on $U\subseteq M$, we
can construct all other kinds of Riemannian metrics on $U$ by means of the
maps $\chi_{p,q}$, $p+q=\dim M$. Note, in the `smooth' case the last
statement may not hold globally on $M$ but it is always valid locally.

	One may ask, what would happen if the signs before the terms in the
r.h.s.\ of~\eref{3.1} are (independently) changed? The change of the sign
before the first term results in the following assertion.

	\begin{Prop}	\label{PropConcl.1}
Let $g$ be a Riemannian metric of signature $(p,q)$ on $U\subseteq M$, $t$ be
a vector field on $U$,
$\widetilde{U}_{\gtreqqless}^{-}:=\{x|\ x\in U,\ -g(t,t)|_x\gtreqqless 1\}$,
and
	\begin{equation}	\label{Concl.1}
g\mapsto \Tilde{g}^- := - g(\cdot,t)\otimes g(\cdot,t) - g.
	\end{equation}
Then $\Tilde{g}^-$ is:

\textup{\textbf{\hphantom{ii}(i)}}
a Riemannian metric with signature $(q,p)$ on $\widetilde{U}_{<}^{-}$.

\textup{\textbf{\hphantom{i}(ii)}}
a Riemannian metric with  signature $(q-1,p+1)$ on $\widetilde{U}_{>}^{-}$.

\textup{\textbf{(iii)}}
a (parabolic) semi-Riemannian metric with signature $(q-1,p)$
and defect $1$ on $\widetilde{U}_{=}^{-}$, \ie on $\widetilde{U}_{=}^{-}$
the bilinear map $h$ has $q$ positive, $(p-1)$ negative, and 1 vanishing
eigenvalue.
	\end{Prop}

	\begin{Proof}
	This proof is almost identical to the one of
proposition~\ref{Prop3.1}. The only difference is that in~\eref{3.h}
$g(t,t)$ must be replaced by $-g(t,t)$ and in~\eref{3.h-isotropic} instead of
$a^2$ must stand $-a^2$. Formally this proof can be obtained from the one of
proposition~\ref{Prop3.1} by replacing in it $t^i$ by $\iu t^i$,
$\iu:=\sqrt{-1}$.
	\end{Proof}

	The change of the sign before the second term in~\eref{3.1} and
in~\eref{Concl.1} is equivalent to put $g=-g^\prime$ with $g^\prime$ being
Riemannian metric with signature $(p,q)$. Then, since $g(t,t)=-g^\prime(t,t)$
and the signature of $g$ is $(q,p)$, we obtain valid versions of
corollary~\ref{Cor3.1} and proposition~\ref{PropConcl.1} if we replace in
them $g$, $p$, and $q$ with $-g$, $q$, and $p$ respectively. Thus we have
proved:

	\begin{Prop}	\label{PropConcl.2}
Let $g$ be a Riemannian metric of signature $(p,q)$ on $U\subseteq M$, $t$ be
a vector field on $U$,
\(
U_{\gtreqqless}^{\pm}	:= \{x|\ x\in U,\ \mp g(t,t)|_x\gtreqqless 1\}
			= \widetilde{U}_{\gtreqqless}^{\mp}
\),
and
	\begin{equation}	\label{Concl.2}
g\mapsto g^\pm := \pm g(\cdot,t)\otimes g(\cdot,t) + g.
	\end{equation}
Then $g^\pm$ is:

\textup{\textbf{\hphantom{ii}(i)}}
a Riemannian metric with signature $(p,q)$ on $U_{<}^{\pm}$.

\textup{\textbf{\hphantom{i}(ii)}}
a Riemannian metric with  signature $(p\pm 1,q\mp 1)$ on $U_{>}^{\pm}$.

\textup{\textbf{(iii)}}
a (parabolic) semi-Riemannian metric with defect $1$ and signature
$(p+(\pm 1 -1)/{2},q+(\mp 1 - 1)/{2})$ on $U_{=}^{\pm}$,
\ie in this case $g^\pm$ has $p+(\pm 1 -1)/{2}$ positive,
$q+(\mp 1 - 1)/{2}$ negative, and 1 vanishing eigenvalue.
	\end{Prop}

	For the metrics $g^\pm$ and $\Tilde{g}^-$ can be proved analogous
results as those for $\Tilde{g}^+:=h$ in sections~\ref{Sect3} and~\ref{Sect4}.
Since this is an almost evident technical task, we do not present them here.
In connection with this, we will note only that the equalities
 $\widetilde{(\Tilde{g}^\pm)}^\pm=g$   and
 $(g^\pm)^\mp=g$
are valid iff
 $\pm g(t,t)=0,+2$ and $\pm g(t,t)=0,-2$ respectively,
while the equations
 $\widetilde{(\Tilde{g}^\pm)}^\mp=g$ and
 $(g^\pm)^\pm=g$
can not be fulfilled for (real) Riemannian metrics as they are equivalent to
 $\pm g(t,t)=1-\iu,1+\iu$ and $\pm g(t,t)=-1-\iu,-1+\iu$ respectively,
$\iu:=+\sqrt{-1}$.

	Metrics like $g^\pm$ find applications in exploring modifications of
general relativity. For instance, up to a positive real constant, the defined
in~\cite[sect.~IV, equation~(4.1)]{Barbero96} metric $g^{\text{Einst}}$ is of
the type $g^\varepsilon$ with
\(
\varepsilon = \sign(-\lambda)
\)
and
\(
t_i = \sqrt{|2\lambda|} \eta_i
\)
with $\lambda:=\frac{\alpha+\beta}{\alpha+2\beta}$,
where
the real parameters $\alpha$ and $\beta$ and the covector $\eta_i$ are
described in~\cite[sect.~II]{Barbero96}.

	A corollary of proposition~\ref{PropConcl.2} is the assertion
of~\cite[sect.~2.6]{Hawking&Ellis}, \cite[p.~219]{Geroch&Horowitz},
and~\cite[p.149, lemma~36]{O'Neill} that if $g$ is an
Euclidean metric and $X$ is non-zero vector field, then
 $h=g-2g(\cdot,X)\otimes g(\cdot,X)/g(X,X)$ is a Lorentzian metric. In fact,
putting  $t=\sqrt{2}X/\sqrt{g(X,X)}$ ($=\sqrt{2}U$ in the notation
of~\cite{O'Neill}), we get $h=g-g(\cdot,t)\otimes g(\cdot,t)$ and
$g(t,t)=-2<-1$.  Therefore $h$ has signature $(n-1,1)$ as that of $g$ is
$(n,0)$, \ie it is a Lorentzian metric according to the accepted
in~\cite{Hawking&Ellis,Geroch&Horowitz,O'Neill} definition.

	There is a simple, but useful for the physics result. Given
metrics $g,\ g^\pm$,  and $\Tilde{g}^\pm$
% with $\Tilde{g}^+:=h$
and a vector field $t$ non\ndash isotropic with respect to $g$
(\ie $g(t,t)\not=0$),
all connected via~\eref{3.1}, \eref{Concl.1}, and~\eref{Concl.2}.
Then there exist (local) fields of bases orthogonal with
respect to all these metrics.
To prove this, we notice that if $\{E_i\}$ is a field of bases with
 $E_n=\lambda t$, $\lambda\not=0,\infty$ and
 $g(E_i,E_j)=\alpha_i\delta_{ij}$, where
 $\alpha_i:M\to\mathbb{R}\backslash\{0\}$ and
 $\delta_{ij}$ are the Kroneker  $\delta$-symbols,
% \ie $\delta_{ij}=0$ for $i\not=j$ and $\delta_{ij}=1$ for $i=j$,
then
 $g^\pm(E_i,E_j) = \beta_{i}^{\pm}\delta_{ij}$,
where
 $\beta_i^\pm = \alpha_i$ for $1\le i<n$ and
 $\beta_n^\pm = \alpha_n \pm \alpha_{n}^{2}/\lambda^2$,
and
% 	$h(E_i,E_j)   = \gamma_{i}^{+}\delta_{ij}$ and
 $\Tilde{g}^\pm(E_i,E_j) = \Tilde{\beta}_{i}^{\pm}\delta_{ij}$
with
 $\Tilde{\beta}_i^\pm = - \alpha_i$ for $1\le i<n$ and
 $\Tilde{\beta}_n^\pm = - \alpha_n \pm \alpha_{n}^{2}/\lambda^2$.

	We end with the remark that the results of this paper may find
possible applications in the theories based on space-time models with
changing signature (topology) (see, \eg~\cite{sign15,sign20}).

\section*{Acknowledgment}

	This work was partially supported by the National Foundation for
Scientific Research of Bulgaria under Grant No.~F642.

\addcontentsline{toc}{section}{References}
\bibliography{bozhopub,bozhoref}
\bibliographystyle{unsrt}
\addcontentsline{toc}{subsubsection}{This article ends at page}
%\addtocontents{toc}{}

\typeout{}
\typeout{}
\typeout{=====================================================================}
\typeout{=====================================================================}
\typeout{}
\typeout{If you want to see some alternative proofs etc.,}
\typeout{remove the first \protect\end{document} command}
\typeout{and run LaTeX twice on the file.}
\typeout{}
\typeout{=====================================================================}
\typeout{=====================================================================}
\typeout{}
\typeout{}

	\end{document}